\documentclass{Interspeech}

\interspeechcameraready

\title{AttentiveMOS: A Lightweight Attention-Only Model for\\Speech Quality Prediction}

\author[affiliation={1}]{Imran E}{Kibria}
\author[affiliation={1}]{Donald S.}{Williamson}

\affiliation{Department of Computer Science \& Engineering}{The Ohio State University}{USA}
\email{kibria.5@osu.edu, williamson.413@osu.com}

\usepackage{comment}
\usepackage{cite}
\usepackage{amsmath,amssymb,amsfonts}
\usepackage{algorithmic}
\usepackage{graphicx}
\usepackage{textcomp}
\usepackage{xcolor}
\usepackage{multirow} 
\usepackage{hyperref}
\usepackage{dblfloatfix}
\usepackage{booktabs}
\usepackage{amsmath}
\usepackage{booktabs}
\usepackage{multirow}

\begin{document}

\maketitle

\begin{abstract}
    

Research in modeling subjective metrics for quality assessment has led to the development of no-reference speech models that directly operate on utterance waveforms to predict the Mean Opinion Score (MOS). These models often rely on convolutional layers for local feature extraction and embeddings from impractically large pretrained networks to enhance generalization. We propose an attention-only model based on Swin transformer and standard transformer layers to extract local context features and global utterance features, respectively. The self-attention operator excels at processing sequences, and our lightweight design enhances generalization on limited MOS datasets while improving real-world applicability. We train our network using a sequential self-teaching strategy to improve generalization on MOS labels affected by noise in listener ratings. Experiments on three datasets confirm the effectiveness of our design and demonstrate improvement over baseline models.

\end{abstract}

\section{Introduction}

Speech quality is a naturalness measure of perceived speech that human listeners best assess. Environmental noises, reverberation, background speakers, networking systems, and signal processing can degrade the quality of perceived speech, since they introduce unwanted sounds and distortions that affect the intended listener. Quality assessment is therefore of significance for the evaluation and improvement of many systems, including those for text-to-speech synthesis, voice conversion, speech separation, and speech enhancement.  

Assessment metrics are classified into objective and subjective categories. Objective metrics are deterministic, whereas subjective metrics depend on human feedback. Early works in deep learning for quality assessment modeled objective metrics, namely Perceptual Evaluation of Speech Quality (PESQ) \cite{pesq} and Short Time Objective Intelligibility (STOI) \cite{stoi}. QualityNet \cite{fu2018quality} was a recurrent model proposed to map the magnitude representations of speech to PESQ. Similarly, the discriminator in MetricGAN \cite{fu2019metricgan} was a convolutional model proposed to map enhanced speech to PESQ or STOI. However, Reddy \textit{et al.} \cite{reddy2022dnsmos} have shown that objective metrics are not strongly correlated with human perception. 

Current research focuses on modeling subjective metrics, primarily the Mean Opinion Score (MOS) \cite{ITU-P800}. It is defined as the averaged listener rating for a given utterance on a scale from 1 (very unnatural) to 5 (completely natural). Many well-performing models predict MOS nonintrusively (requiring only degraded input), since a clean speech reference is not available in many real-world scenarios \cite{dong2020towards}. Deep Noise Suppression Mean Opinion Score (DNSMOS) \cite{reddy2022dnsmos} is a convolutional model, MOSNet \cite{lo2019mosnet} and Listener-Dependent Network (LDNet) \cite{huang2022ldnet} are hybrid convolutional recurrent models, while NISQA \cite{mittag2021nisqa} is a hybrid convolutional attention model, all proposed to map only the magnitude representations of degraded speech to MOS. 

Previous models used only speech magnitude representations as input, as the phase is unstructured and has historically been considered unimportant for a long time \cite{wang1982unimportance}. However, studies have now shown the importance of phase in denoising \cite{zhang2020investigation}\cite{williamson2015complex}, establishing the waveform input (implicitly containing both magnitude and phase) as more informational. UTMOS \cite{saeki2022utmos} and SSL-MOS \cite{cooper2022generalization} are large models based on self-supervised learning (SSL) proposed to map an utterance waveform to the MOS metric. Their SSL component, wav2vec \cite{baevski2020wav2vec} and HuBERT \cite{hsu2021hubert} models, use a stack of 1D convolutional layers to generate features directly from an utterance waveform. Similarly, MTQNet \cite{zezario2024multi} uses SSL features, but concatenates power spectral features and learnable filter banks to predict MOS. In general, SSL models achieve high accuracy but are prohibitively large (e.g., $\sim$95M parameters), making them unsuitable for resource-constrained applications like earbuds or hearing aids, which typically run on microcontrollers with less than 1\,MB of RAM and require inference latency under 10\,ms.


Inspired by the success of attention-only architectures in NLP \cite{vaswani2017attention}, vision \cite{dosovitskiy2020image}, and speech \cite{gong2021ast}, we propose AttentiveMOS. It is an attention-only architecture (most optimally suited to sequential inputs) to model the subjective MOS metric using an utterance waveform as the input. The network size, consisting of only 86K parameters, is significantly lighter than the SSL-based quality prediction models. A smaller network size facilitates generalization in sparingly available MOS datasets and precludes the need for pre-training or transfer learning. Our AttentiveMOS architecture is a cascade of Swin transformer \cite{liu2021swin} and standard transformer layers. The acoustic properties of an utterance vary over time. We propose using a Swin transformer to locally capture these temporally varying acoustic features. The Swin transformer was originally proposed to operate on windows of image patches having spatial structure; however, we adapt it to groups of temporal frames of speech (termed contexts). The local context features are further aggregated to global utterance features using standard transformers. Our training strategy aims to improve generalization under subjective variance in labels. This subjective variance arises from the fact that multiple listeners have different ratings for the same utterance signal. We treat it as noise in the true MOS label and use a multistage self-teaching strategy proposed in \cite{kumar2020sequential}. Finally, experiments with the Samsung Open Mean Opinion Score (SOMOS) \cite{maniati2022somos} and the Blizzard Voice Conversion Challenge (BVCC) \cite{huang2022voicemos} datasets validate our design and show improvements over existing baselines on error and correlation metrics.

\begin{figure*}[!h]
\centering
\includegraphics[width=\textwidth]{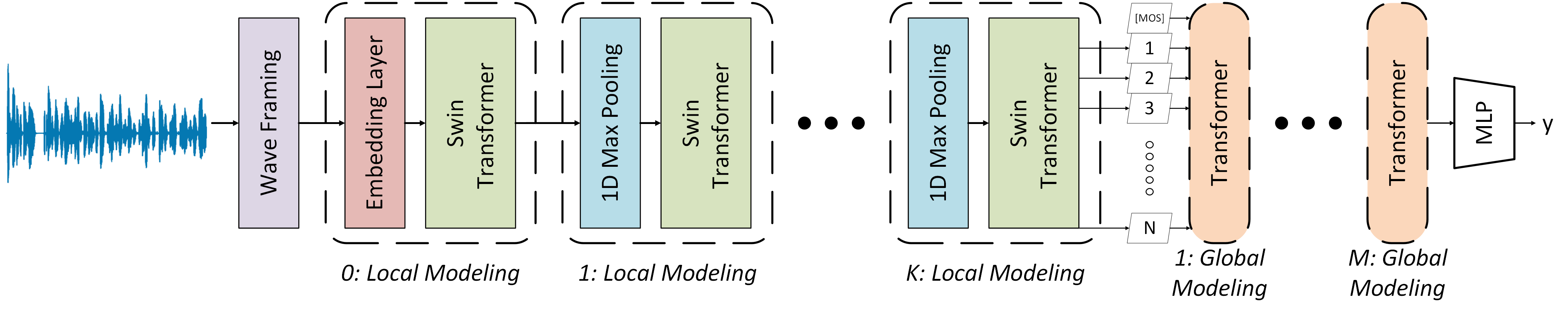}
\caption{The architectural diagram of our proposed AttentiveMOS model. $y$ corresponds to the MOS prediction of the input utterance. }
\label{fig1}
\end{figure*}

\section{Proposed Approach}

In the following subsections, we elaborate on our model design, its underlying intuition, and its training strategy. Our training aims at improving generalization when the MOS ratings of an utterance vary between multiple listeners.

\subsection{Model Architecture}\label{AA}

The block diagram of our architectural design is shown in Fig.~\ref{fig1}. In the first step, a 1D waveform $\mathbb{R}^n$ of duration $T$ seconds is reshaped into small frames of $2ms$ with an overlap of $1ms$. This operation is performed in the wave framing block, the output of which is a 2D representation $\mathbb{R}^{F \times S}$ of frames $F$ and samples per frame $S$ for a single utterance. The frame size is chosen small because it directly correlates with the model size. 

In the second step, a linear embedding layer maps each frame to a latent space of dimension $D$, hence the output $\mathbb{R}^{F \times D}$. Typically, positional information is added to feature embeddings before feeding to a transformer. This is because the self-attention operation in the transformer is permutation invariant. However, we did not add any positional encodings to our frame embeddings. We experimentally observed that it deteriorated the performance of our model. We believe that modeling the sequential structure is relevant to learning linguistic content in speech, but perceptual quality content (including noise) is independent of sequential dependencies. 

In the third step, a Swin transformer captures local dependencies within groups of frame tokens. This contrasts with a standard transformer that captures global dependencies across all input tokens. The acoustic features vary across an utterance, which motivates our use of Swin transformer to capture local dependencies. We further delineate its operation in subsection \ref{subsec:swin_transformer}.  Its output has the same dimensions $\mathbb{R}^{F \times D}$ as the input, but encompasses local dependencies within groups of frame tokens. We term a group of successive frame tokens \textit{context}. 

In the fourth step, the output of the Swin transformer is fed to a max pooling layer. The 1D max pooling operation merges $m_{i}$ frame tokens in each block $i$ to compress information along the time axis. For example, the output dimensions of block $i$$=$$1$ will be $\mathbb{R}^{(F/m_{1}) \times D}$. Our choice of max pooling differs from the proposition in \cite{liu2021swin}, where a linear layer was used to merge the image patches. However, our experiments show that the max pooling performs better than a linear layer. We believe it is also because a linear layer applied to consecutive frame embeddings attempts to model sequential structure, whereas acoustic content is independent of such sequential dependencies. We call an embedding or max-pooling layer together with a Swin transformer a \textit{local modeling} block. A cascade of $K+1$ such blocks reduces the number of frames and captures local dependencies within contexts, resulting in $\mathbb{R}^{N \times D}$ dimensional output. 

In the fifth step, we prepend a learnable [MOS] token at the start of the $N$ frame tokens. The $\mathbb{R}^{(1+N) \times D}$ dimensional input is fed to a standard transformer, also termed a \textit{global modeling} block. This standard transformer applies attention across all tokens, hence aggregating acoustic information for the whole utterance. This information is also shared with the [MOS] token embeddings during self-attention. In total, we have $M$ global modeling blocks cascaded sequentially, resulting in $\mathbb{R}^{(1+N) \times D}$ dimensional output.

In the final step, we have a shallow multilayer perceptron mapping the $\mathbb{R}^{1 \times D}$ embeddings of the [MOS] token to a scalar value. This value is the MOS prediction corresponding to the input utterance. Our network has 86K trainable parameters in total when the embedding size $D$$=$$16$. 

\subsection{Swin Transformer} \label{subsec:swin_transformer}

\begin{figure}[!t]
    \centering
    \includegraphics[width=1.\columnwidth]{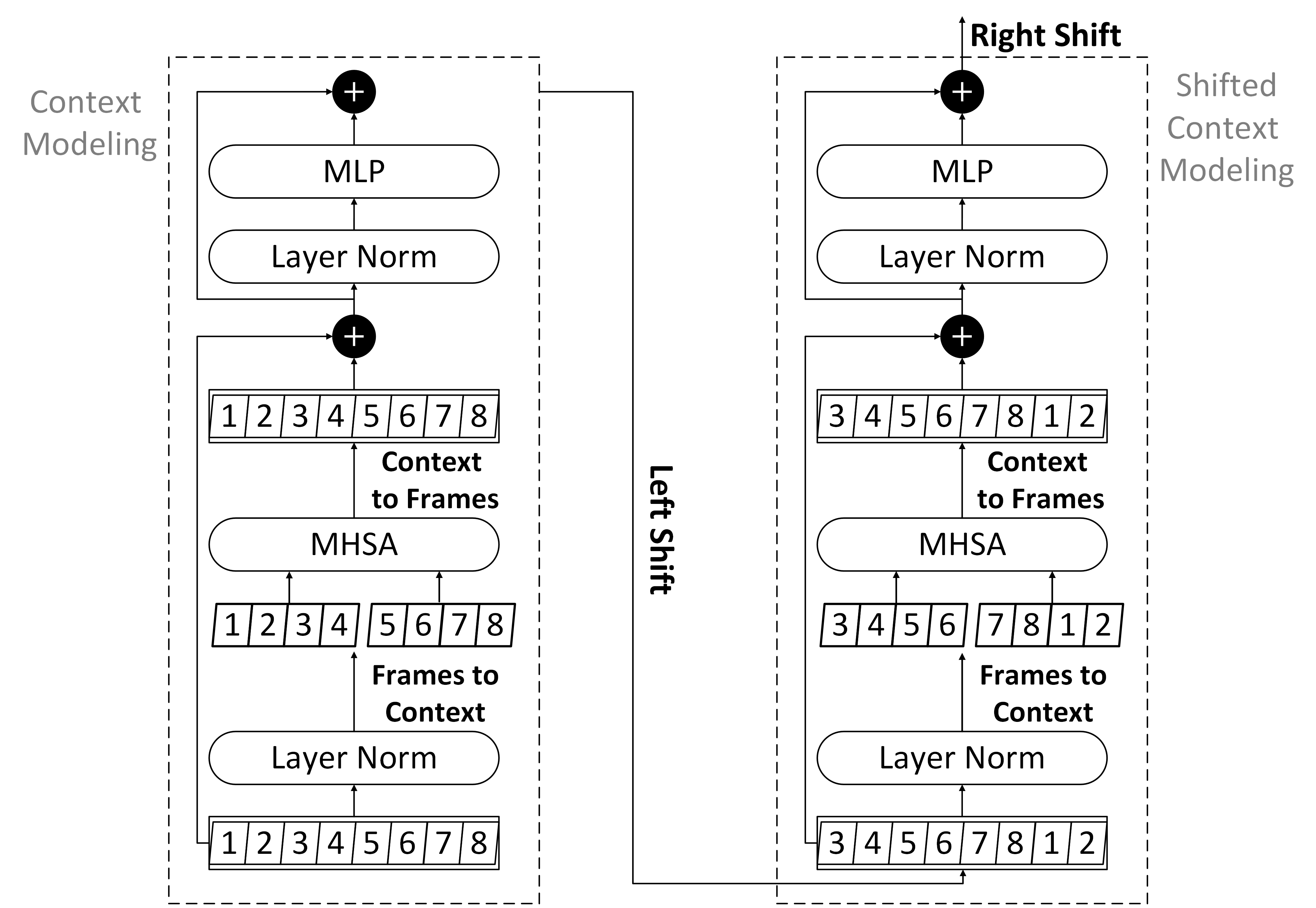}
    \caption{Swin transformer for speech with 8 frame tokens.}
    \label{fig2}
\end{figure}

The Swin transformer \cite{liu2021swin} was originally proposed for images; however, we adapted it for speech. We explain its mechanism using a toy example of an utterance with eight frame tokens. The architecture diagram is shown in Fig.~\ref{fig2}. It consists of two standard transformer layers stacked one on top of the other. The first layer models dependencies within contexts, whereas the second layer models dependencies within shifted contexts.  Shifted contexts ensure that dependencies at context disjunctions in the first layer are also captured.

In the first transformer layer (left part in Fig.~\ref{fig2}), a layer normalization step normalizes the features of all frame tokens in the utterance. Next, a frame-to-context operation groups frame tokens into disjoint contexts via a reshaping operator, so only local rather than global dependencies are captured. Multihead self-attention (MHSA) captures these local dependencies within each context. Following this, a context-to-frame operation reverts contexts to frame tokens via reshaping. A residual term is added to all frame tokens through a skip connection. This result is then passed through a sequence of layer normalization (LN), multilayer perceptron (MLP), and another skip connection to yield the output of the first transformer layer.

In the second transformer layer (the right portion in Fig.~\ref{fig2}), the input is circularly shifted to the left by half the context size. This shifting ensures that dependencies at disjunctions of previous contexts are now captured in the newly formed shifted contexts. The architecture of this transformer layer is identical to the first one, except for masking during self-attention. It can be seen that the start tokens $[1,2]$ and the end tokens $[7,8]$ are within the same context in the shifted sequence $[7,8,1,2]$. We want to prevent modeling the cross-interaction between the start and end tokens because they do not occur together in the original sequence. For this reason, we use a boolean mask to allow only end tokens $[7,8]$ to interact mutually and start tokens $[1,2]$ to interact mutually but prevent their cross-interaction.

\subsection{Training Objective}

The mean opinion score (MOS) label is the average of multiple listener ratings collected for an utterance signal. We hypothesize that MOS labels of all utterances are not an equally good estimate of their subjective quality. Utterances with a small standard deviation in their listener ratings have strong MOS labels, whereas utterances with a large standard deviation in their listener ratings have weak MOS labels. Therefore, we propose penalizing an utterance prediction based on its distance from the label and the quality of that label. Our training objective is given as follows.
\begin{equation}
    l = \ln{(1 + \frac{|y - \mu|}{\sigma + \epsilon})}\label{eq}
\end{equation}
where $y$ is our model prediction, $\mu$ is the MOS label, $\sigma$ is the standard deviation in listener ratings, and $\epsilon$ is a small constant empirically determined at $0.01$.

\subsection{Sequential Learning}

The sequential self-teaching framework (SUSTAIN) \cite{kumar2020sequential} was proposed to iteratively improve the generalization performance of a supervised model on noisy labeled datasets. It involves training a sequence of models, wherein the model in the current stage receives co-supervision from the model in the previous stage(s). This co-supervision is derived from a convex combination of the dataset label and the outputs of one or more models from earlier stages. At any stage, teacher-student models can have an identical architecture, hence the name self-teaching. 

We treat the difference in listener ratings of the same utterance as noise in an unknown true label and adopt the SUSTAIN framework as in \cite{reddy2022dnsmos}. A series of models using the AttentiveMOS architecture are trained to enhance generalization despite the subjective variance in MOS labels. The training label for co-supervision at any stage is determined as follows.  
\begin{equation}
    \mu_{m} = (\alpha_{0} \cdot \mu) + \sum_{i=1}^{m} (\alpha_{i} \cdot y_{i-1})
    \label{eq:3}
\end{equation}
where $\mu_m$ and $\mu$ correspond to the student label in stage $m$ and the MOS label in the dataset, respectively. $y_{i-1}$ corresponds to the prediction of the model from previous stages. The data set label and teacher predictions are weighted by $\alpha_{i}$, where the sum of all weights $\alpha_{i}$ must equal 1. 

\section{Experimentation}

\subsection{Datasets}

In this study, we used three MOS labeled datasets, namely the Samsung Open Mean Opinion Score (SOMOS) \cite{maniati2022somos} from the speech synthesis domain, the Blizzard Voice Conversion Challenge (BVCC) \cite{huang2022voicemos} from the voice conversion domain, and the Non-Intrusive Speech Quality Assessment (NISQA) \cite{yi2022conferencingspeech} from the speech enhancement domain. A statistical summary of these datasets is given in Table \ref{tab:dataset}, where $N_{\text{min}}$ is the number of minimum of ratings per sample and $\sigma_{\text{avg}}$ is the standard deviation in averaged across training set.

\begin{table}[t!]
    \caption{Statistical Summary of Datasets.}

    \centering
    \begin{tabular}{lcccc c}
        \toprule
        \multirow{2}{*}{Dataset} & \multicolumn{3}{c}{\# Samples} & \multirow{2}{*}{$N_{\text{min}}$} & \multirow{2}{*}{$\sigma_{\text{avg}}$} \\
        \cmidrule(lr){2-4}
        & Train & Dev & Test &  &  \\
        \midrule
        SOMOS  & 14,000 & 3,000 & 3,000 & 17 & 1.124 \\
        BVCC & 4,974 & 1,066 & 1,066 & 8 & -- \\
        NISQA  & 11,020 & 2,700 & 952 & 5 & 0.666 \\
        \bottomrule
    \end{tabular}

    \label{tab:dataset}
\end{table}

The SOMOS dataset contains utterances synthesized from neural text-to-speech systems in a single-speaker scenario. It consists of two tracks based on label quality, namely SOMOS-full and SOMOS-clean. The SOMOS-full dataset contains all listener ratings for each utterance, while the SOMOS-clean dataset has been manually filtered to remove spurious ratings. 


\subsection{Experimental Setup}

All utterances are sampled at 16kHz. The duration of all input utterances is set at $20.48s$, signals shorter than this duration are padded with trailing silence. The wave framing block divides utterances into frames of $2ms$ with $1ms$ overlap. A small frame requires a small token, which leads to a lightweight network, since the token size $D$ is directly correlated with the number of trainable parameters.

AttentiveMOS has $K$$+$$1$ local modeling blocks, where $K=6$. The size of the context in each Swin transformer is [$10$, $4$, $4$, $4$, $4$, $2$, $2$]. This context size in a Swin transformer is analogous to the kernel size in a convolutional layer. The kernel size in the max-pooling layers is [$5$, $2$, $2$, $2$, $2$, $2$]. Local modeling blocks reduce $F=20480$ frames to $N=128$ tokens. Subsequently, the $M=12$ standard transformers apply attention to all tokens to generate global features. Lastly, a shallow MLP having two dense layers with GELU activation followed by a linear layer maps the \textit{[MOS]} token embeddings to a scaler value. All hyperparameter values are determined experimentally and from the literature.

For training, we used AdamW optimizer with a fixed $10^{-4}$ learning rate and gradient clipping. The batch size was set at $8$, a small number for good regularization, and the model is trained separately on each dataset for $250$ epochs. 

\subsection{Architecture and Loss Analysis}

The design decisions for the architecture are analyzed on the SOMOS dataset in Table \ref{tab:pe_fm_loss}. These decisions pertain to positional encodings and merging frames in local modeling blocks. For positional encodings, validation performance of two architecture configurations is compared. In one configuration, fixed sinusoidal encodings are added to frame embeddings before providing the input to the Swin transformer, whereas the other configuration excludes positional encodings. The performance with learnable encodings is not analyzed because it significantly increases the network size, making a fair comparison infeasible. Our results show that including positional information deteriorates validation performance. For merging frames, validation performance of three architecture configurations is compared. In one configuration, a linear layer inputs a group of frame tokens (equivalent to kernel size in pooling layers) and merges them to a single frame token. Alternatively, an average-pooling and a max-pooling layer are used in the other two configurations, respectively. Our results show that max pooling yields the best performance. These results follow that acoustic features in an utterance are independent of sequential structuring.


Next, we compare the validation performance of AttentiveMOS on the SOMOS dataset when it is optimized for three different loss functions: Mean Absolute Error (MAE), Mean Squared Error (MSE), and our proposed loss function \eqref{eq}. MAE and MSE penalize a model prediction only based on its distance from the MOS label, whereas our proposed loss function penalizes a prediction based on both the distance from MOS label and its subjective quality. 
We observe that the MSE objective results in the best performance and that our proposed objective does not perform as well. Perhaps it is because the error and correlation metrics assign equal importance to all (prediction, label) pairs. In reality, each label is not an equally strong quality representation of its utterance. We will further explore this idea in future work. 

\begin{table}[t]
    \centering
    \caption{Analysis of Positional Encodings (PE), Frame Merging (FM), \& Training Loss using SOMOS Validation Set.}
    \begin{tabular}{@{}lcccccc@{}}
        \toprule
        & \multicolumn{2}{c}{\textbf{Clean Set}} & \multicolumn{2}{c}{\textbf{Full Set}} \\
        \cmidrule(lr){2-3} \cmidrule(lr){4-5} 
        \textbf{} & \textbf{MSE}$\downarrow$ & \textbf{PCC}$\uparrow$ & \textbf{MSE}$\downarrow$ & \textbf{PCC}$\uparrow$ \\
        \midrule
        PE: Included  & 0.292 & 0.344 & 0.140 & 0.318 \\
        PE: Excluded  & \textbf{0.265} & \textbf{0.422} & \textbf{0.132} & \textbf{0.394} \\
        \midrule
        FM: Linear    & 0.304 & 0.289 & 0.148 & 0.256 \\
        FM: Avg Pool  & 0.292 & 0.347 & 0.138 & 0.336 \\
        FM: Max Pool   & \textbf{0.265} & \textbf{0.422} & \textbf{0.132} & \textbf{0.394} \\
        \midrule
        Loss: MAE     & 0.267 & 0.408 & 0.128 & 0.397 \\
        Loss: MSE     & \textbf{0.257} & \textbf{0.449} & \textbf{0.127} & \textbf{0.412} \\
        Loss: Ours    & 0.265 & 0.422 & 0.132 & 0.394 \\
        \bottomrule
    \end{tabular}
    \label{tab:pe_fm_loss}
\end{table}

\subsection{Baseline Comparison}

A performance comparison of AttentiveMOS with baseline models is given in Table \ref{tab:utterance-level}. The results are reported on the SOMOS-clean validation set when each network has separately been trained on three datasets. The three training datasets, SOMOS-clean, SOMOS-full, and BVCC are in-distribution, out-of-distribution, and out-of-domain with the SOMOS-clean validation set, respectively. The baseline models include MOSNet \cite{lo2019mosnet}, LDNet \cite{huang2022ldnet} and SSL-MOS \cite{cooper2022generalization}, all of which are fairly recent and frequently used in the literature. Their reported performance is taken from \cite{maniati2022somos}, where researchers trained each model from scratch. It is important that all models be trained from scratch on the same dataset for a fair evaluation of the proposed architecture, since a model performance is subject to data availability and the use of pretrained embeddings.   


 Our results show that AttentiveMOS has better performance than MOSNet and LDNet, and comparable performance to SSL-MOS when trained on the SOMOS-clean set. However, SSL-MOS is based on the wav2vec architecture, which has 95M parameters, whereas our model has only 86K, making it significantly more lightweight. AttentiveMOS achieves the best performance on all metrics when trained on the SOMOS-full set. However, it does not generalize well when trained on the BVCC dataset which is out-of-domain with the evaluation set. 

We analyzed AttentiveMOS on the NISQA dataset. It achieved PCC values of $0.861$, $0.769$, $0.666$, $0.558$, and $0.359$ on simulated training set, live training set, P501 test set, FOR test set, and LIVETALK test set (out-of-domain), respectively. 

\begin{table}[t!]
    \centering
    \caption{Models separately trained on three datasets and evaluated on the SOMOS-clean Validation set.}
    \begin{tabular}{@{}lcccccc@{}}
    \toprule
        \textbf{Model (trainset)} & \textbf{MSE}$\downarrow$ & \textbf{PCC}$\uparrow$ & \textbf{SRCC}$\uparrow$ \\
        \midrule
        MOSNet \cite{lo2019mosnet} (VCC2018)      & 0.843 & -0.075 & -0.084 \\
        MOSNet \cite{lo2019mosnet} (SOMOS-full)   & 0.598 &  0.218 &  0.238 \\
        MOSNet \cite{lo2019mosnet} (SOMOS-clean)  & 0.729 &  0.352 &  0.347 \\
        \midrule
        LDNet \cite{huang2022ldnet} (BVCC)          & 1.011 &  0.040 &  0.032 \\
        LDNet \cite{huang2022ldnet} (SOMOS-full)    & 0.581 &  0.262 &  0.275 \\
        LDNet \cite{huang2022ldnet} (SOMOS-clean)   & 0.642 &  0.397 &  0.386 \\
        \midrule
        SSL-MOS \cite{cooper2022generalization} (BVCC)        & 2.217 & \textbf{0.229} & \textbf{0.230} \\
        SSL-MOS \cite{cooper2022generalization} (SOMOS-full)  & 0.564 &  0.296 &  0.313 \\
        SSL-MOS \cite{cooper2022generalization} (SOMOS-clean) & 0.625 &  \textbf{0.453} &  \textbf{0.444} \\
        \midrule
        AttentiveMOS (BVCC)           & \textbf{0.390} & 0.113 & 0.109 \\
        AttentiveMOS (SOMOS-full)     & \textbf{0.312} & \textbf{0.445} &  \textbf{0.433} \\
        AttentiveMOS (SOMOS-clean)    & \textbf{0.257} &  0.449 &  0.442 \\
        \bottomrule
    \end{tabular}
    \label{tab:utterance-level}
\end{table}

\subsection{Impact of Sequential Learning}

The generalization performance with sequential learning on MOS labels is shown in Table \ref{tab3}. The results are reported on the SOMOS-full set because it includes spurious ratings from listeners in addition to authentic ones, hence the generalization over noisy labels can be better observed. The MSE loss is used for training in all stages since teacher predictions are deterministic, hence zero variance in labels. The $\alpha$ values in stages $m=1$, $m=2$, and $m=3$ are [$0.4$, $0.6$], [$0.3$, $0.3$, $0.4$], and [$0.225$, $0.225$, $0.25$, $0.3$] respectively. We observed student model in later stages converge quicker and show better performance. The performance begins to saturate after $m=3$.

\begin{table}[htbp]
    \centering
    \caption{Self-teaching Performance on SOMOS-Full Set}

    \begin{tabular}{@{}lcccccc@{}}
    \toprule
    \multirow{2}{*}{\textbf{Stage}} & \multicolumn{2}{c}{\textbf{Validation Set}} & \multicolumn{2}{c}{\textbf{Test Set}} \\
    \cmidrule(lr){2-3} \cmidrule(lr){4-5}
     & \textbf{MSE}$\downarrow$ & \textbf{PCC}$\uparrow$ 
     & \textbf{MSE}$\downarrow$ & \textbf{PCC}$\uparrow$ \\
    \midrule

    Base   & 0.127 & 0.412 & 0.126 & 0.403 \\
    $m=1$  & 0.126 & 0.409 & 0.123 & 0.414 \\
    $m=2$  & \textbf{0.123} & \textbf{0.431} & \textbf{0.121} & \textbf{0.432} \\
    $m=3$  & 0.124 & 0.428 & 0.121 & 0.429 \\
    
    \bottomrule
    \end{tabular}

    \label{tab3}
\end{table}

\section{Conclusion}

We proposed a lightweight attention-only network for modeling speech quality by capturing local context and global utterance features. Our network improves on existing baselines and sequential learning improves generalization on MOS labels affected by noise in listener ratings. 

\bibliographystyle{IEEEtran}
\bibliography{mybib}

\begin{thebibliography}{10}
\providecommand{\url}[1]{#1}
\csname url@samestyle\endcsname
\providecommand{\newblock}{\relax}
\providecommand{\bibinfo}[2]{#2}
\providecommand{\BIBentrySTDinterwordspacing}{\spaceskip=0pt\relax}
\providecommand{\BIBentryALTinterwordstretchfactor}{4}
\providecommand{\BIBentryALTinterwordspacing}{\spaceskip=\fontdimen2\font plus
\BIBentryALTinterwordstretchfactor\fontdimen3\font minus \fontdimen4\font\relax}
\providecommand{\BIBforeignlanguage}[2]{{%
\expandafter\ifx\csname l@#1\endcsname\relax
\typeout{** WARNING: IEEEtran.bst: No hyphenation pattern has been}%
\typeout{** loaded for the language `#1'. Using the pattern for}%
\typeout{** the default language instead.}%
\else
\language=\csname l@#1\endcsname
\fi
#2}}
\providecommand{\BIBdecl}{\relax}
\BIBdecl

\bibitem{pesq}
A.~W. Rix, J.~G. Beerends, M.~P. Hollier, and A.~P. Hekstra, ``Perceptual evaluation of speech quality (pesq) - a new method for speech quality assessment of telephone networks and codecs,'' in \emph{Proc. IEEE International Conference on Acoustics, Speech and Signal Processing}, 2001, pp. 749--752.

\bibitem{stoi}
C.~H. Taal, R.~C. Hendriks, R.~Heusdens, and J.~Jensen, ``An algorithm for intelligibility prediction of time--frequency weighted noisy speech,'' \emph{IEEE Transactions on Audio, Speech, and Language Processing}, pp. 2125--2136, 2011.

\bibitem{fu2018quality}
S.-W. Fu, Y.~Tsao, H.-T. Hwang, and H.~Wang, ``Quality-net: An end-to-end non-intrusive speech quality assessment model based on blstm,'' in \emph{Proc. ISCA Interspeech}, 2018, pp. 1873--1877.

\bibitem{fu2019metricgan}
S.-W. Fu, C.-F. Liao, Y.~Tsao, and S.-D. Lin, ``Metricgan: Generative adversarial networks based black-box metric scores optimization for speech enhancement,'' in \emph{Proc. PMLR International Conference on Machine Learning}, 2019, pp. 2031--2041.

\bibitem{reddy2022dnsmos}
C.~K. Reddy, V.~Gopal, and R.~Cutler, ``Dnsmos p. 835: A non-intrusive perceptual objective speech quality metric to evaluate noise suppressors,'' in \emph{Proc. IEEE International Conference on Acoustics, Speech and Signal Processing}, 2022, pp. 886--890.

\bibitem{ITU-P800}
\BIBentryALTinterwordspacing
``{ITU-T Recommendation P.800: Methods for subjective determination of transmission quality},'' International Telecommunication Union, Tech. Rep., February 1998. [Online]. Available: \url{https://www.itu.int/rec/T-REC-P.800}
\BIBentrySTDinterwordspacing

\bibitem{dong2020towards}
X.~Dong and D.~S. Williamson, ``Towards real-world objective speech quality and intelligibility assessment using speech-enhancement residuals and convolutional long short-term memory networks,'' \emph{AIP The Journal of the Acoustical Society of America}, pp. 3348--3359, 2020.

\bibitem{lo2019mosnet}
C.-C. Lo, S.-W. Fu, W.-C. Huang \emph{et~al.}, ``Mosnet: Deep learning based objective assessment for voice conversion,'' in \emph{Proc. ISCA Interspeech}, 2019, pp. 1541--1545.

\bibitem{huang2022ldnet}
W.-C. Huang, E.~Cooper, J.~Yamagishi, and T.~Toda, ``Ldnet: Unified listener dependent modeling in mos prediction for synthetic speech,'' in \emph{Proc. IEEE International Conference on Acoustics, Speech and Signal Processing}, 2022, pp. 896--900.

\bibitem{mittag2021nisqa}
G.~Mittag, B.~Naderi, A.~Chehadi, and S.~M{\"o}ller, ``Nisqa: A deep cnn-self-attention model for multidimensional speech quality prediction with crowdsourced datasets,'' in \emph{Proc. ISCA Interspeech}, 2021, pp. 2127--2131.

\bibitem{wang1982unimportance}
D.~Wang and J.~Lim, ``The unimportance of phase in speech enhancement,'' \emph{IEEE Transactions on Acoustics, Speech, and Signal Processing}, pp. 679--681, 1982.

\bibitem{zhang2020investigation}
Z.~Zhang, D.~S. Williamson, and Y.~Shen, ``Investigation of phase distortion on perceived speech quality for hearing-impaired listeners,'' in \emph{Proc. ISCA Interspeech}, 2020, pp. 2512--2516.

\bibitem{williamson2015complex}
D.~S. Williamson, Y.~Wang, and D.~Wang, ``Complex ratio masking for monaural speech separation,'' \emph{IEEE Transactions on Audio, Speech, and Language Processing}, pp. 483--492, 2015.

\bibitem{saeki2022utmos}
T.~Saeki, D.~Xin, W.~Nakata \emph{et~al.}, ``Utmos: Utokyo-sarulab system for voicemos challenge 2022,'' in \emph{Proc. ISCA Interspeech}, 2022, pp. 4521--4525.

\bibitem{cooper2022generalization}
E.~Cooper, W.-C. Huang, T.~Toda, and J.~Yamagishi, ``Generalization ability of mos prediction networks,'' in \emph{Proc. IEEE International Conference on Acoustics, Speech and Signal Processing}, 2022, pp. 8442--8446.

\bibitem{baevski2020wav2vec}
A.~Baevski, Y.~Zhou, A.~Mohamed \emph{et~al.}, ``wav2vec 2.0: A framework for self-supervised learning of speech representations,'' \emph{Advances in neural information processing systems}, pp. 12\,449--12\,460, 2020.

\bibitem{hsu2021hubert}
W.-N. Hsu, B.~Bolte, Y.-H.~H. Tsai \emph{et~al.}, ``Hubert: Self-supervised speech representation learning by masked prediction of hidden units,'' \emph{IEEE/ACM transactions on audio, speech, and language processing}, pp. 3451--3460, 2021.

\bibitem{zezario2024multi}
R.~E. Zezario, B.-R.~B. Bai, C.-S. Fuh \emph{et~al.}, ``Multi-task pseudo-label learning for non-intrusive speech quality assessment model,'' in \emph{Proc. IEEE International Conference on Acoustics, Speech and Signal Processing}, 2024, pp. 831--835.

\bibitem{vaswani2017attention}
A.~Vaswani, N.~Shazeer, N.~Parmar \emph{et~al.}, ``Attention is all you need,'' in \emph{Advances in Neural Information Processing Systems}, 2017.

\bibitem{dosovitskiy2020image}
A.~Dosovitskiy, L.~Beyer, A.~Kolesnikov \emph{et~al.}, ``An image is worth 16x16 words: Transformers for image recognition at scale,'' in \emph{International Conference on Learning Representations}, 2020.

\bibitem{gong2021ast}
Y.~Gong, Y.-A. Chung, and J.~R. Glass, ``Ast: Audio spectrogram transformer,'' in \emph{Proc. ISCA Interspeech}, 2021, pp. 571--575.

\bibitem{liu2021swin}
Z.~Liu, Y.~Lin, Y.~Cao \emph{et~al.}, ``Swin transformer: Hierarchical vision transformer using shifted windows,'' in \emph{Proc. IEEE International Conference on Computer Vision}, 2021, pp. 10\,012--10\,022.

\bibitem{kumar2020sequential}
A.~Kumar and V.~Ithapu, ``A sequential self teaching approach for improving generalization in sound event recognition,'' in \emph{Proc. PMLR International Conference on Machine Learning}, 2020, pp. 5447--5457.

\bibitem{maniati2022somos}
G.~Maniati, A.~Vioni, N.~Ellinas \emph{et~al.}, ``Somos: The samsung open mos dataset for the evaluation of neural text-to-speech synthesis,'' in \emph{Proc. ISCA Interspeech}, 2022, pp. 2388--2392.

\bibitem{huang2022voicemos}
W.-C. Huang, E.~Cooper, Y.~Tsao \emph{et~al.}, ``The voicemos challenge 2022,'' in \emph{Proc. ISCA Interspeech}, 2022, pp. 4536--4540.

\bibitem{yi2022conferencingspeech}
G.~Yi, W.~Xiao, Y.~Xiao \emph{et~al.}, ``Conferencingspeech 2022 challenge: Non-intrusive objective speech quality assessment (nisqa) challenge for online conferencing applications,'' in \emph{Proc. ISCA Interspeech}, 2022, pp. 3308--3312.

\end{thebibliography}

\end{document}